\documentclass{iau}
\usepackage{graphicx}

\title[High-resolution spectropolarimetric of Kappa Cet] %% give here short title %%
{High-resolution spectropolarimetric of {$\kappa$ Cet}: A proxy for the young Sun}

\author[do Nascimento et al.]   %% give here short author list %%
{J.D. do Nascimento Jr$^1$, 
P. Petit$^2$, 
S.  Marsden$^3$, 
G. F.  Porto de Mello$^4$, 
I. Ribas$^5$, 
S. Jeffers$^6$, 
M. Castro$^{1}$, 
E. Guinan$^{7}$\\ \and the  Bcool Collaboration}

\affiliation{
$^{1}$DFTE, Univ. Federal do R. Grande do Norte, 59072-970 Natal, RN, Brazil  email: {\tt dias@dfte.ufrn.br}\\
$^{2}$ IRAP, CNRS,  14 Av.  Edouard Belin, F-31400 Toulouse, France\\
$^{3}$ CESRC, Univ. of Southern Queensland, Toowoomba, 4350, Australia\\
$^{4}$OV, UFRJ, L. do Pedro Ant\^onio, 43 20080-090 Rio de Janeiro, RJ, Brazil\\
$^{5}$IES  de Catalunya,  UAB, F. de Ci\`{e}ncies, Torre, 08193 Bellaterra, Spain\\
$^{6}$IAG, Georg-August-Univ.  G\"{o}ttingen, Friedrich-Hund-Platz 1, 37077 G\"{o}ttingen, Germany\\
$^{7}$Department of Astronomy and Astrophysics, Villanova University, Villanova, PA 19085; US}

\pubyear{2013}
\volume{302}  %% insert here IAU Symposium No.
\pagerange{119--126}
\jname{Magnetic Fields Throughout Stellar Evolution}
\editors{P. Petit, M.M. Jardine \& H.C. Spruit, eds.}
\begin{document}

\maketitle

\begin{abstract}

\end{abstract}
Among the solar proxies studied in the Sun in Time  {$\kappa^{\rm 1}$ Cet~}(HD 20630) stands out as potentially having a mass very close to solar and a young age.  On this study, we monitored the magnetic field and the chromospheric activity from the Ca II H \& K lines of {$\kappa^{\rm 1}$ Cet}.  We used the Least-Square-Deconvolution (LSD, \cite[Donati et al. 1997]{donati97}) by simultaneously extracting the information contained in all 8,000 photospheric lines of the echelogram (K1  type star). To reconstruct a reliable magnetic map and characterize the surface differential rotation of {$\kappa^{\rm 1}$ Cet~}  we used 14 exposures spread over 2 months, in order to cover at least two rotational cycles (Prot $\sim$ 9.2 days). The LSD technique was applied to detect the Zeeman signature of the magnetic field in each of our 14 observations and to measure its longitudinal component. In order to reconstruct the magnetic field geometry of {$\kappa^{\rm 1}$ Cet~}, we applied the Zeeman Doppler Imaging (ZDI) inversion method. ZDI revealed a structure in the radial magnetic field consisting of a polar magnetic spot. On this study, we present the fisrt look results of a high-resolution spectropolarimetric campaign to characterize the activity and the magnetic fields of this young solar proxy. 

\firstsection % if your document starts with a section,

\section{Introduction}
\label{introduction}

The  observational programme  {\it Sun in Time}   is focused on a small 
sample of carefully-selected and well-studied stellar proxies that  well  represent some  key stages in the evolution 
of the Sun.  Among the 
solar proxies  studied in the Sun in Time,  {$\kappa^{\rm 1}$ Cet~}   stands out as potentially having a  mass and metallicity
 very  close to solar with an estimated age of $\sim$ 0.7 Gyr  (\cite[Ribas et al. 2005]{ribas05}). This could be a very good analog of the 
Sun at the critical time when life is thought to have  originated on Earth 3.8 Gyr ago. The star was discovered to have a rapid rotation, 
roughly once every nine days. {$\kappa^{\rm 1}$ Cet~}  is also considered a  good candidate to contain terrestrial planets.
In spite of our in-depth knowledge of {$\kappa^{\rm 1}$ Cet~}, including its radiative properties, abundances, atmospheric parameters, and 
evolutionary state, we know little  or nothing about the magnetic field properties  of this important  solar proxy.  
On this study we present the first look results from a regularly  observational campaign using the NARVAL spectropolarimeter at the
T\'elescope Bernard Lyot (Pic du Midi, France) to observe {$\kappa^{\rm 1}$ Cet~},  and infer the intensity and nature of its magnetic field.  The Zeeman-Doppler Imaging technique is employed to reconstruct the large-scale photospheric magnetic field structure of {$\kappa^{\rm 1}$ Cet~}  and to investigate its short-term temporal evolution. 

\section{First look results}
\label{results}

From a careful analysis, \cite[Ribas et al. (2010)]{ribas10} compare  different methods and give
for {$\kappa^{\rm 1}$ Cet~}  the following atmospheric parameters: Teff = 5665 $\pm$ 30 K, log g = 4.49 $\pm$ 0.05 dex   and [Fe/H] = +0.10 ± 0.05.  In the left panel of Fig. 1, we show the normalized Stokes V profiles 
of {$\kappa^{\rm 1}$ Cet~}. 
\begin{figure}[h]
% \vspace*{-2.0 cm}
\begin{center}
  \includegraphics[width=1.3in]{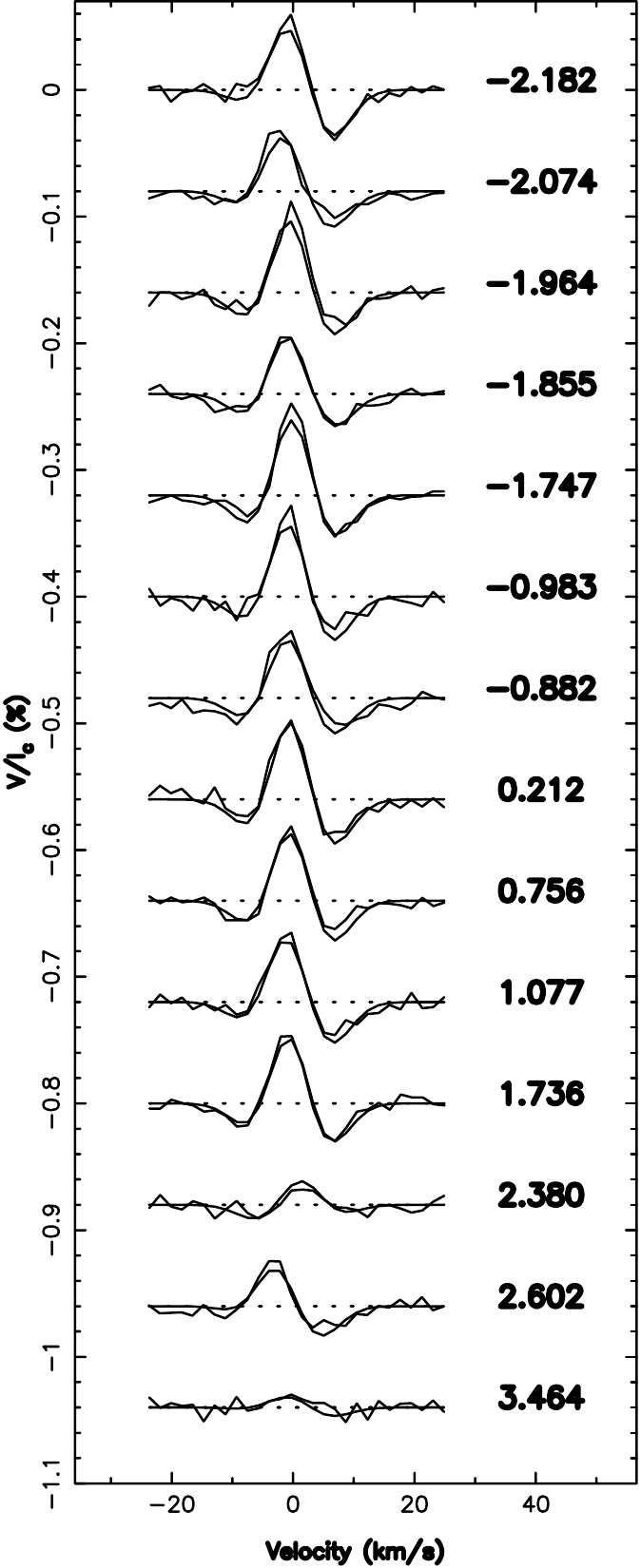}
  \qquad
  \includegraphics[width=3in]{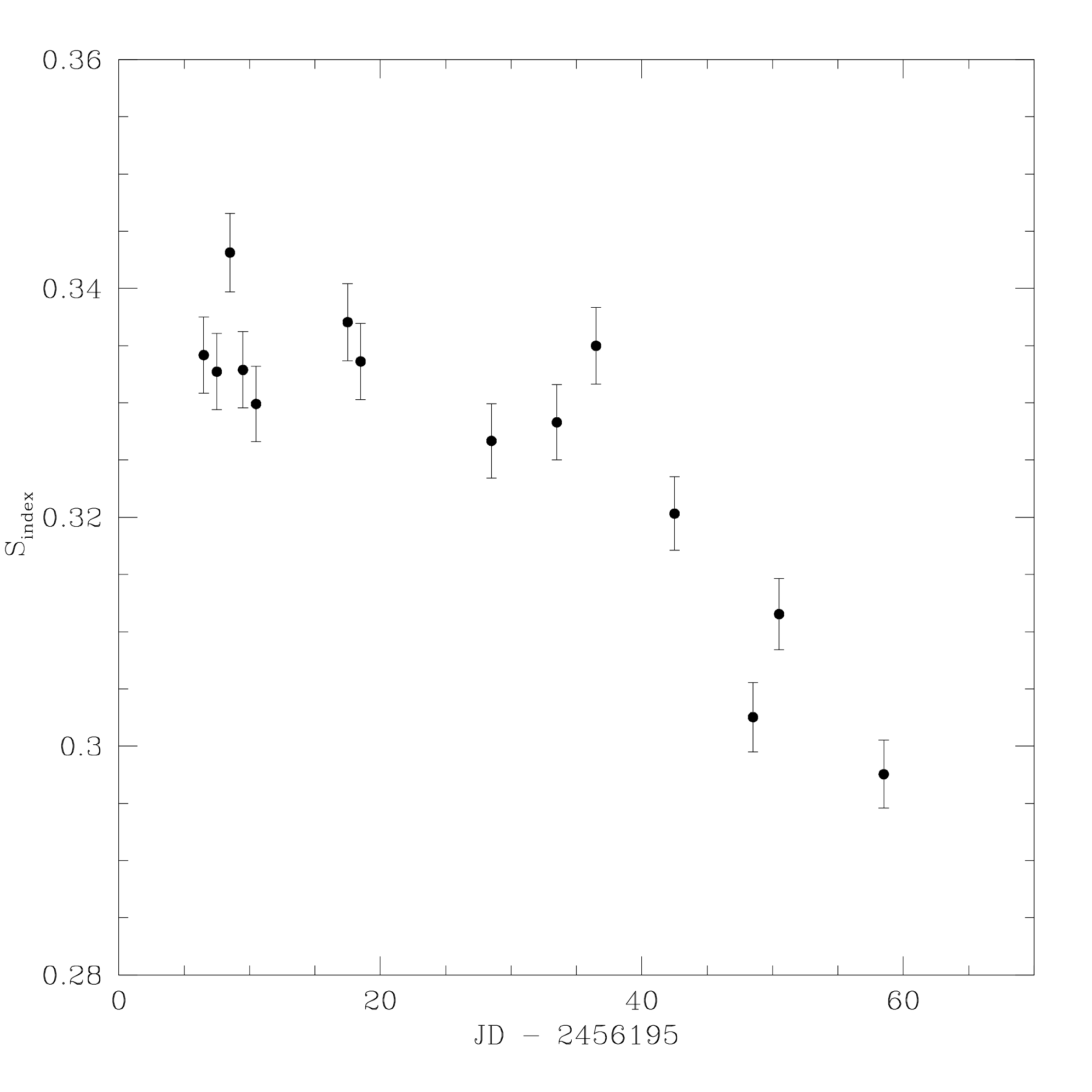} 
% \vspace*{-1.0 cm}
 \caption{\textit{Left}: The normalized Stokes V profiles  of {$\kappa^{\rm 1}$ Cet~}.  Continuum line represent the data and dashed line correspond to synthetic profiles of our magnetic model.  Rotational phases of observations are indicated in the right part of the plot. \textit{Right}: From the Stokes I spectra, we determined the S$_{index}$  to quantify the changes in the Ca II H line. The pipeline is described in \cite[Morgenthaler et al. (2012)]{morgenthaler12} and \cite[Wright et al. (2004)]{wright04}.}
   \label{logtauc&Ro_BV}
\end{center}
\end{figure}
The LSD technique was applied to detect the Zeeman signature of the magnetic field in each of our 14 observations to measure its longitudinal component, leading to B$_{l}$~=+7.0~$\pm$0.9 Gauss. The main structure observed in the ZDI model is a structure in the radial magnetic field consisting of a polar magnetic spot. In this study, we present the first look results of a high-resolution spectropolarimetric campaign to characterize the activity and the magnetic fields of this young solar proxy.

\end{document}